# Observation of a Narrow State Decaying into $\Xi_c^+\pi^-$

P. Avery,[1] A. Freyberger,[1] K. Lingel,[1] C. Prescott,[1] J. Rodriguez,[1] S. Yang,[1] J. Yelton,[1] G. Brandenburg,[2] D. Cinabro,[2] T. Liu,[2] M. Saulnier,[2] R. Wilson,[2] H. Yamamoto,[2] T. Bergfeld,[3] B.I. Eisenstein,[3] J. Ernst,[3] G.E. Gladding,[3] G.D. Gollin,[3] M. Palmer,[3] M. Selen,[3] J.J. Thaler,[3] K.W. Edwards,[4] K.W. McLean,[4] M. Ogg,[4] A. Bellerive,[5] D.I. Britton,[5] E.R.F. Hyatt,[5] R. Janicek,[5] D.B. MacFarlane,[5] P.M. Patel,[5] B. Spaan,[5] A.J. Sadoff,[6] R. Ammar,[7] P. Baringer,[7] A. Bean,[7] D. Besson,[7] D. Coppage,[7] N. Copty,[7] R. Davis,[7] N. Hancock,[7] S. Kotov,[7] I. Kravchenko,[7] N. Kwak,[7] Y. Kubota,[8] M. Lattery,[8] M. Momayezi,[8] J.K. Nelson,[8] S. Patton,[8] R. Poling,[8] V. Savinov,[8] S. Schrenk,[8] R. Wang,[8] M.S. Alam,[9] I.J. Kim,[9] Z. Ling,[9] A.H. Mahmood,[9] J.J. O'Neill,[9] H. Severini,[9] C.R. Sun,[9] S. Timm,[9] F. Wappler,[9] G. Crawford,[10] J.E. Duboscq,[10] R. Fulton,[10] D. Fujino,[10] K.K. Gan,[10] K. Honscheid,[10] H. Kagan,[10] R. Kass,[10] J. Lee,[10] M. Sung,[10] C. White,[10] A. Wolf,[10] M.M. Zoeller,[10] X. Fu,[11] B. Nemati,[11] W.R. Ross,[11] P. Skubic,[11] M. Wood,[11] M. Bishai,[12] J. Fast,[12] E. Gerndt,[12] J.W. Hinson,[12] T. Miao,[12] D.H. Miller,[12] M. Modesitt,[12] E.I. Shibata,[12] I.P.J. Shipsey,[12] P.N. Wang,[12] L. Gibbons,[13] S.D. Johnson,[13] Y. Kwon,[13] S. Roberts,[13] E.H. Thorndike,[13] T.E. Coan,[14] J. Dominick,[14] V. Fadeyev,[14] I. Korolkov,[14] M. Lambrecht,[14] S. Sanghera,[14] V. Shelkov,[14] T. Skwarnicki,[14] R. Stroynowski,[14] I. Volobouev,[14] G. Wei,[14] M. Artuso,[15] M. Gao,[15] M. Goldberg,[15] D. He,[15] N. Horwitz,[15] S. Kopp,[15] G.C. Moneti,[15] R. Mountain,[15] F. Muheim,[15] Y. Mukhin,[15] S. Playfer,[15] S. Stone,[15] X. Xing,[15] J. Bartelt,[16] S.E. Csorna,[16] V. Jain,[16] S. Marka,[16] D. Gibaut,[17] K. Kinoshita,[17] P. Pomianowski,[17] B. Barish,[18] M. Chadha,[18] S. Chan,[18] D.F. Cowen,[18] G. Eigen,[18] J.S. Miller,[18] C. O'Grady,[18] J. Urheim,[18] A.J. Weinstein,[18] F. Würthwein,[18] D.M. Asner,[19] M. Athanas,[19] D.W. Bliss,[19] W.S. Brower,[19] G. Masek,[19] H.P. Paar,[19] J. Gronberg,[20] C.M. Korte,[20] R. Kutschke,[20] S. Menary,[20] R.J. Morrison,[20] S. Nakanishi,[20] H.N. Nelson,[20] T.K. Nelson,[20] C. Qiao,[20] J.D. Richman,[20] D. Roberts,[20] A. Ryd,[20] H. Tajima,[20] M.S. Witherell,[20] R. Balest,[21] K. Cho,[21] W.T. Ford,[21] M. Lohner,[21] H. Park,[21] P. Rankin,[21] J.G. Smith,[21] J.P. Alexander,[22] C. Bebek,[22] B.E. Berger,[22] K. Berkelman,[22] K. Bloom,[22] T.E. Browder,[22*] D.G. Cassel,[22] H.A. Cho,[22] D.M. Coffman,[22] D.S. Crowcroft,[22] M. Dickson,[22] P.S. Drell,[22] D.J. Dumas,[22] R. Ehrlich,[22] R. Elia,[22] P. Gaidarev,[22] M. Garcia-Sciveres,[22] B. Gittelman,[22] S.W. Gray,[22] D.L. Hartill,[22] B.K. Heltsley,[22] S. Henderson,[22] C.D. Jones,[22] S.L. Jones,[22] J. Kandaswamy,[22] N. Katayama,[22] P.C. Kim,[22] D.L. Kreinick,[22] T. Lee,[22] Y. Liu,[22] G.S. Ludwig,[22] J. Masui,[22] J. Mevissen,[22] N.B. Mistry,[22] C.R. Ng,[22] E. Nordberg,[22] J.R. Patterson,[22] D. Peterson,[22] D. Riley,[22] and A. Soffer[22]

(CLEO Collaboration)

[1] University of Florida, Gainesville, Florida 32611
[2] Harvard University, Cambridge, Massachusetts 02138
[3] University of Illinois, Champaign-Urbana, Illinois, 61801
[4] Carleton University, Ottawa, Ontario K1S 5B6 and the Institute of Particle Physics, Canada
[5] McGill University, Montréal, Québec H3A 2T8 and the Institute of Particle Physics, Canada
[6] Ithaca College, Ithaca, New York 14850
[7] University of Kansas, Lawrence, Kansas 66045
[8] University of Minnesota, Minneapolis, Minnesota 55455
[9] State University of New York at Albany, Albany, New York 12222
[10] Ohio State University, Columbus, Ohio, 43210
[11] University of Oklahoma, Norman, Oklahoma 73019
[12] Purdue University, West Lafayette, Indiana 47907
[13] University of Rochester, Rochester, New York 14627
[14] Southern Methodist University, Dallas, Texas 75275
[15] Syracuse University, Syracuse, New York 13244
[16] Vanderbilt University, Nashville, Tennessee 37235
[17] Virginia Polytechnic Institute and State University, Blacksburg, Virginia, 24061
[18] California Institute of Technology, Pasadena, California 91125
[19] University of California, San Diego, La Jolla, California 92093
[20] University of California, Santa Barbara, California 93106
[21] University of Colorado, Boulder, Colorado 80309-0390
[22] Cornell University, Ithaca, New York 14853

## Abstract

Using data recorded by the CLEO-II detector at CESR, we report the first observation of a narrow state decaying into $\Xi_c^+\pi^-$. The state has mass difference $M(\Xi_c^+\pi^-) - M(\Xi_c^+)$ of $178.2 \pm 0.5 \pm 1.0$ MeV/c$^2$, and a width of $< 5.5$ MeV/c$^2$ (90% confidence level limit). The most likely explanation of this new state is that it is the $J = \frac{3}{2}$ spin excitation of the $\Xi_c^0$ charmed baryon.

*Permanent address: University of Hawaii at Manoa

Earlier CLEO [1,2] and other experimental groups [3–6] have reported the observation of a ground state isodoublet, the $\Xi_c^+$ and $\Xi_c^0$ charmed baryons ($J^P = \frac{1}{2}^+$) . In these states the two lighter quarks are antisymmetric under interchange of flavor (i.e. in a spin-0 configuration). The next highest states are expected to be the $J^P = \frac{1}{2}^+$ $\Xi_c'$ and the $J^P = \frac{3}{2}^+$ $\Xi_c^*$ states, in which the lighter quarks are symmetric under the exchange of flavor and thus in a spin-1 configuration. According to theoretical predictions [7–11] the masses of the $\Xi_c'$ states are expected to be below threshold for the decay to $\Xi_c\pi$, in which case they will decay electromagnetically; there has been one preliminary result indicating a possible signal in $\Xi_c'^+ \to \Xi_c^+\gamma$ with a mass difference of around 95 MeV/c$^2$ [12]. On the other hand, the $\Xi_c^*$ states are expected to be heavy enough to decay by emission of a $\pi^\pm$. In this Letter, we present evidence of the existence of a particle decaying into $\Xi_c^+\pi^-$. In view of the theoretical models for the mass spectrum, we identify this state as the $\Xi_c^{*0}$.

The data presented here were taken by the CLEO II detector operating at the Cornell Electron Storage Ring. The sample used in this analysis corresponds to an integrated luminosity of 3.7 $fb^{-1}$ from data taken on the $\Upsilon(4S)$ resonance and in the continuum at energies just above and below the $\Upsilon(4S)$.

The CLEO II detector is described elsewhere [13]; here we will briefly describe the parts of the detector most relevent to this analysis. The CLEO II detector is designed to detect both charged and neutral particles with excellent resolution and efficiency. The detector consists of a charged particle tracking system surrounded by a scintillation counter time-of-flight system and an electromagnetic shower detector consisting of 7800 thallium-doped cesium iodide crystals. These detectors are installed within a 1.5T superconducting solenoidal magnet. Particle identification is achieved by a combination of time-of-flight measurements and of energy-loss measurements in the drift chamber. In this analysis tracks are assigned a particular hypothesis if they are have measurements loosely consistent with that particle; the efficiency of this requirement is around 99% per charged track. One of the modes studied includes a $K^-$, and for this $K^-$ a more restrictive identification requirement; we require the probability the candidate is a $K$ to be at least 50% of the sum of the probabilities for the

$\pi$, $K$, and $p$ hypotheses.

We report the observation of a new particle decaying into $\Xi_c^+\pi^-$, where the $\Xi_c^+$ charmed baryon has been observed decaying into either $\Xi^-\pi^+\pi^+$ ($\Xi^- \to \Lambda\pi^-, \Lambda \to p\pi^-$), $\Xi^0\pi^+\pi^0$ ($\Xi^0 \to \Lambda\pi^0, \Lambda \to p\pi^-$), or $\Sigma^+K^{*0}$ ($\Sigma^+ \to p\pi^0, K^{*0} \to K^-\pi^+$).* These decay modes of the $\Xi_c^+$ were chosen because they have the most significant signals. We have presented measurements [14,15] of the relative branching fraction of the $\Xi_c^+$ decaying into these channels. The analysis presented here is similar to that of Reference 14, but optimized for greater detection efficiency, and includes an augmented data set.

Candidates for $\Lambda$ decays are reconstructed from pairs of oppositely charged tracks, assuming the higher momentum one to be the proton and the lower momentum one to be the pion. The measured momentum components are propagated through the magnetic field to the candidate decay vertex, which is required to be greater than 2 mm from the primary event vertex. The candidates are required to have a measured invariant mass within 5.0 MeV/c$^2$ ($\approx 3\sigma$) of the known $\Lambda$ mass. The tracks were then kinematically fit to this mass and used to reconstruct $\Xi$'s.

The $\Xi^-$ candidates were formed by combining each $\Lambda$ candidate with each remaining negatively charged track. A vertex is formed from the intersection of the $\Lambda$ track and the negatively charged track. The momentum components of the charged track are recalculated at the candidate $\Xi^-$ vertex. We require that the measured flight path of the reconstructed $\Xi^-$ be greater than 2 mm, the reconstructed $\Xi^-$ be consistent with coming from the main event vertex, and the measured distance between the event vertex and the $\Xi^-$ decay point be less than the distance between the event vertex and the $\Lambda$ vertex. Combinations with a measured invariant mass within 5 MeV/c$^2$ ($\approx 3\sigma$) of the known $\Xi^-$ mass were kinematically fit to this mass and used to reconstruct $\Xi_c^+$ candidates.

Candidates for $\Xi^0$ baryons were formed by combining each $\Lambda$ candidate with each $\pi^0$

*Charge conjugate modes are implicit throughout.

candidate. These $\pi^0$ candidates were formed from a pair of photons detected in the CsI calorimeter. As a first approximation they are assumed to come from the event vertex and only a loose cut is applied on the $\gamma\gamma$ invariant mass. The $\Lambda$ candidates used for $\Xi^0$ reconstruction were required to have a measured flight path of greater than 1.5 cm and to not point back to the event vertex. The $\Xi^0$ is assumed to be created at the event vertex, and to have a momentum equal to the sum of the momenta of the $\Lambda$ and $\pi^0$ candidates. The decay point of the $\Xi^0$ is taken to be the point of intersection between the $\Xi^0$ candidate and the $\Lambda$ candidate. This decay point was required to be at least 3 mm from the event vertex. The 4-momentum of the $\pi^0$ candidate was recalculated using the $\Xi^0$ decay point as the point of origin of the photons, and its mass was required to be consistent ($< 3.5\sigma$) with the known $\pi^0$ mass. The $\Lambda\pi^0$ invariant mass was then recalculated using this improved estimate of the $\pi^0$ momentum, and those combinations within 8 MeV/c$^2$ ($\approx 3\sigma$) of the known $\Xi^0$ mass were kinematically fit to this mass and used to reconstruct $\Xi_c^+$'s.

Tracks were defined to be candidates for protons from $\Sigma^+$ decays if their energy loss and time-of-flight were consistent with a proton hypothesis, their momenta is greater than 0.5 GeV, and their impact parameter with respect to the event vertex in the plane perpendicular to the beam direction is more than 0.6 mm. $\Sigma^+$ candidates were then constructed from $p\pi^0$ combinations and a $\Sigma^+$ decay point found in a manner similar to the $\Xi^0$ decay point described above. The reconstructed decay vertex position was required to be consistent with arising from a positive lifetime decay. Those $p\pi^0$ combinations within 15 MeV/c$^2$ ($\approx 3\sigma$) of the known $\Sigma^+$ mass were kinematically constrained to this mass and used to reconstruct $\Xi_c^+$ candidates.

In order to select $\Xi_c^+$ candidates, each $\Xi^-$ was combined with each remaining $\pi^+\pi^+$ pair in the event and each $\Xi^0$ was combined with each remaining $\pi^+\pi^0$ pair, where these $\pi^0$ candidates were required to have $p > 300$ MeV/c to reduce the background to the signal. The $\Sigma^+$ candidates were combined with $K^-\pi^+$ combinations and the reconstructed $K^-\pi^+$ invariant mass was required to be within 50 MeV/c$^2$ of the $K^{*0}$ mass. In order to reduce the combinatorial background, which is worst for $\Xi_c^+$ candidates with low momentum, we apply a mode-dependent cut on $x_p$, where $x_p = p/p_{max}$; $p$ is the momentum of the charmed baryon, $p_{max} = \sqrt{E_{beam}^2 - M^2}$, and $E_{beam}$ is the beam energy. Charmed baryons produced from decays of B mesons are kinematically limited to $x_p < 0.4$, so this cut rejects those candidates, leaving only those produced by $e^+e^-$ annihilation into $c\bar{c}$ jets, which are known to have a hard momentum spectrum. To illustrate the good signal to noise ratio of the $\Xi_c^+$ signal, the invariant mass spectrum of $\Xi^-\pi^+\pi^+$ combinations with $x_p > 0.4$ is shown in Figure 1(a); Figure 1(b) shows the spectrum of $\Xi^0\pi^+\pi^0$ combinations with $x_p > 0.6$, and Figure 1(c) shows the spectrum of $\Sigma^+K^{*0}$ combinations with $x_p > 0.5$. In the fits, which are overlayed on these figures, the signals are paramaterized by Gaussians with fixed widths ($\sigma$ = 7 MeV/c$^2$,$\sigma$ = 15 MeV/c$^2$ and $\sigma$ = 9 MeV/c$^2$, respectively); they show yields of $160 \pm 18$, $76 \pm 12$, and $59 \pm 12$ events. These widths were determined using a GEANT based Monte Carlo simulation of the detector. The background functions used were polynomials, and in Figure 1(c) there is an added background due to the reflection of misidentified $\Lambda_c^+ \to \Sigma^+\pi^+\pi^-$ events. Combinations within $2.5\sigma$ of the mass of the $\Xi_c^+$ in each decay mode are taken as $\Xi_c^+$ candidates. The $x_p$ cut used in Figure 1 was applied only for the purposes of illustrating the quality of these signals and was released before continuing with the analysis; we prefer to apply an $x_p$ cut only on the $\Xi_c^+\pi^-$ combination.

The $\Xi_c^+$ candidates defined above were then combined with each remaining $\pi^-$ track and the mass difference $M(\Xi_c^+\pi^-) - M(\Xi_c^+)$ was calculated. We then placed an $x_p$ cut on the $\Xi_c^+\pi^-$ combination, $x_p > 0.4$ for those involving the decay $\Xi^-\pi^+\pi^+$, $x_p > 0.6$ for those involving $\Xi^0\pi^+\pi^0$, and $x_p > 0.5$ for those involving $\Sigma^+K^{*0}$. The mass difference plot, shown in Figure 2, shows a clear peak at around 178 MeV/c$^2$ . We fit this mass spectrum to the sum of a Chebychev polynomial with threshold suppression, and a Breit-Wigner convoluted with a Gaussian resolution function ($\sigma$ = 1.6 MeV/c$^2$ , calculated by Monte Carlo program). The fit yields a signal area of $54.6 \pm 12.1$ combinations, a mean mass difference of $178.2 \pm 0.5$ MeV/c$^2$ , and an intrinsic width, $\Gamma = 2.6^{+1.7}_{-1.4}$ MeV/c$^2$ , where the errors shown are statistical errors only. Considering systematic errors due to the fitting procedures and to energy-loss corrections for charged tracks, we find a mass difference for this new state of $178.2 \pm 0.5 \pm 1.0$ MeV/c$^2$ . The measurement of the width is consistent with zero, so we present a 90% confidence level upper limit of $\Gamma < 5.5$ MeV/c$^2$ .

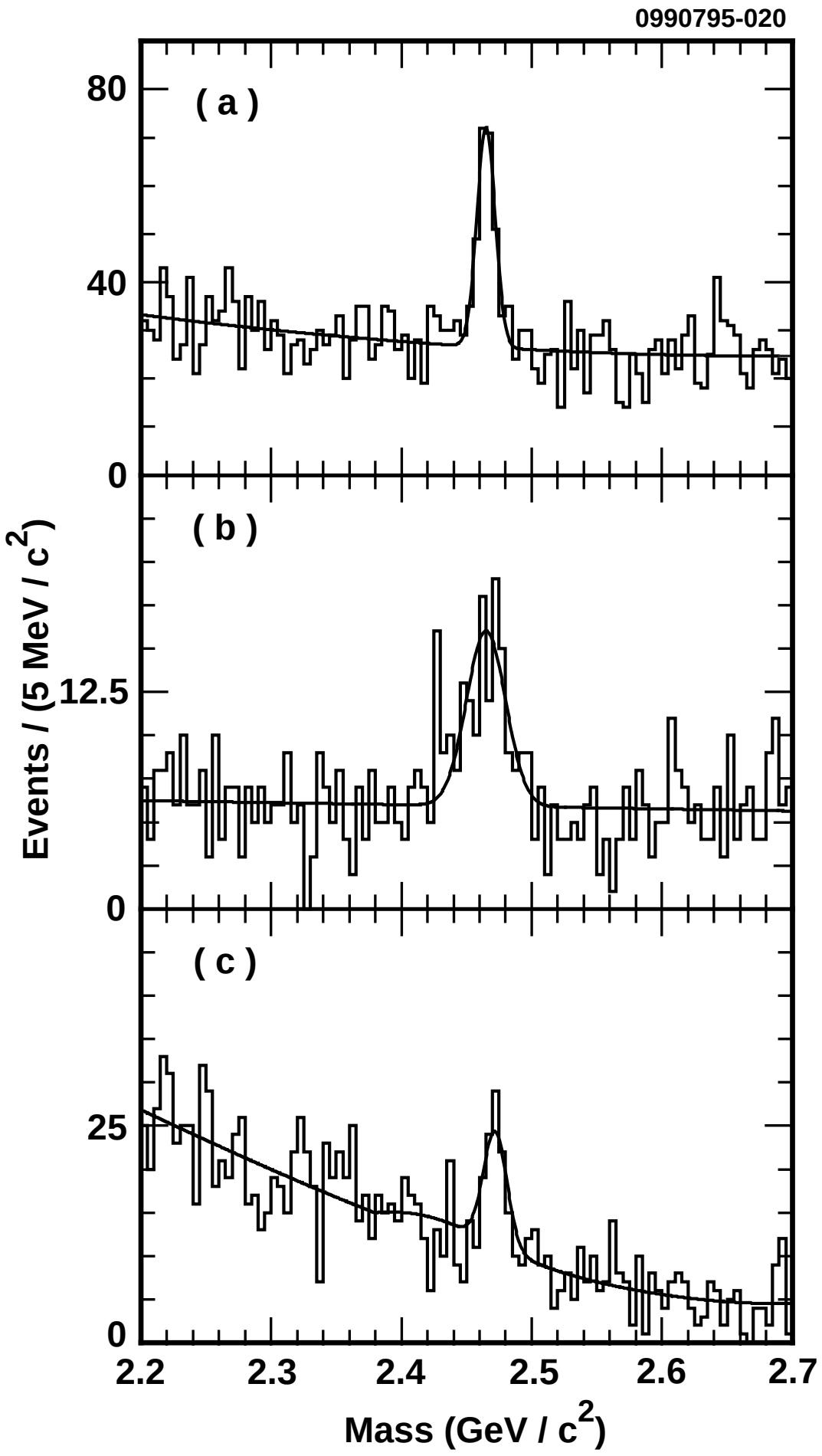

FIG. 1. Combinations of (a) $\Xi^-\pi^+\pi^+$ with $x_p > 0.4$, (b) $\Xi^0\pi^+\pi^0$ with $x_p > 0.6$, and (c) $\Sigma^+K^{*0}$ with $x_p > 0.5$. All show clear $\Xi_c^+$ peaks. The fits are described in the text.

Figures 3(a), 3(b), and 3(c), respectively, show the same mass difference as presented in Figure 2, but separated into combinations involving the three $\Xi_c^+$ decay chains separately. In the fits overlayed on these histograms, the mass and width of the signal were constrained to the values found by the fit to Figure 2. The number of events in the peaks are found to be $31.8 \pm 6.6$ events for Figure 3(a), $10.5 \pm 4.6$ events for Figure 3(b), and $10.9 \pm 4.3$ for Figure 3(c).

We identify this new state as the $\Xi_c^{*0}$. In order to study the fragmentation function we study only those events in which $\Xi_c^+ \to \Xi^-\pi^-\pi^+$, as this mode has a good signal as low as $x_p = 0.4$. We divide the data shown in Figure 3(a) into bins of $x_p$ from 0.4 to 1.0, determine the $\Xi_c^{*0}$ yield in each bin, and correct the yields using efficiencies obtained from Monte Carlo calculations. Figure 4 shows $\frac{1}{N}\frac{dN}{dx}$, for data points from $x_p = 0.4$ to $x_p = 1.0$. The overlayed fit, which is to the form of the fragmentation function of Peterson et al. [16], gives an $\epsilon = 0.22^{+.15}_{-.08}$. This is similar to that obtained for $\Xi_c^+$ baryons [17], but larger than that of the $L = 1$ charmed baryons [18]. Using all three decay chains, and extrapolating the efficiency-corrected $\Xi_c^+$ and $\Xi_c^{*0}$ baryons yields down to $x_p = 0$, we calculate that $(27 \pm 6 \pm 6)\%$ of the $\Xi_c^+$'s come from $\Xi_c^{*0}$ decays, where the uncertainties are statistical and systematic respectively.

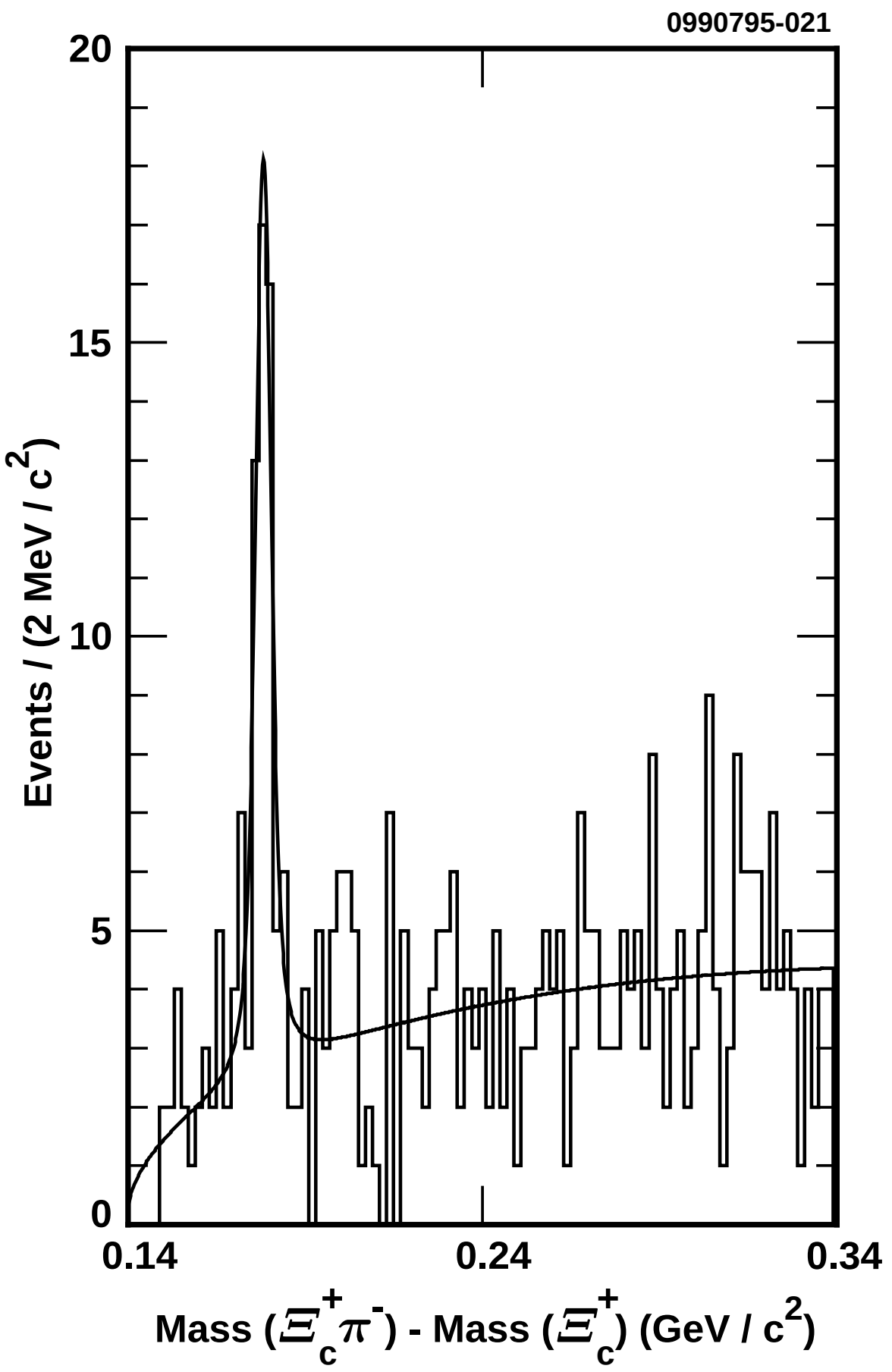

FIG. 2. The spectrum of the mass difference $M(\Xi_c^+\pi^-) - M(\Xi_c^+)$ for all three decay chains.

Our identification of the new state as the $J = \frac{3}{2}^+$ state relies upon theoretical models. Taking the mass difference above and adding the $\Xi_c^+$ mass of $2465.1 \pm 1.6$ MeV/c$^2$ [19], we obtain a $\Xi_c^{*0}$ mass of $2642.8 \pm 2.2$ MeV/c$^2$ . The model predictions for this state are in the range 2620 to 2690 MeV/c$^2$ [7–11]. Our measurement is not consistent with the expectations for the $\Xi_c^{\prime 0}$ state by the same authors, nor is it similar to the preliminary measurement of the $\Xi_c^{\prime +}$ state reported by WA-89 [12]. Orbital (L=1) excitations of $\Xi_c$ states would be expected to occur at higher mass differences, as they do in the the $\Lambda_c^+$ system [18]. The expected width of a $J = \frac{3}{2}^+$ state can be calculated by analogy with the non-charmed $\Xi^{*0}$. We expect $\Gamma(\Xi_c^{*0})/\Gamma(\Xi^{*0})$ to be $0.75 p_1^3/p_2^3$ where $p_1$ and $p_2$ are the decay momenta for the two processes, and where 0.75 is the appropriate ratio of the overlap of the spin wave-functions [20]. Using our measured value of the mass difference this calculation leads to a expected width of the $\Xi_c^{*0}$ of $\approx 2.5$ MeV/c$^2$ , consistent with our observation of a narrow state.

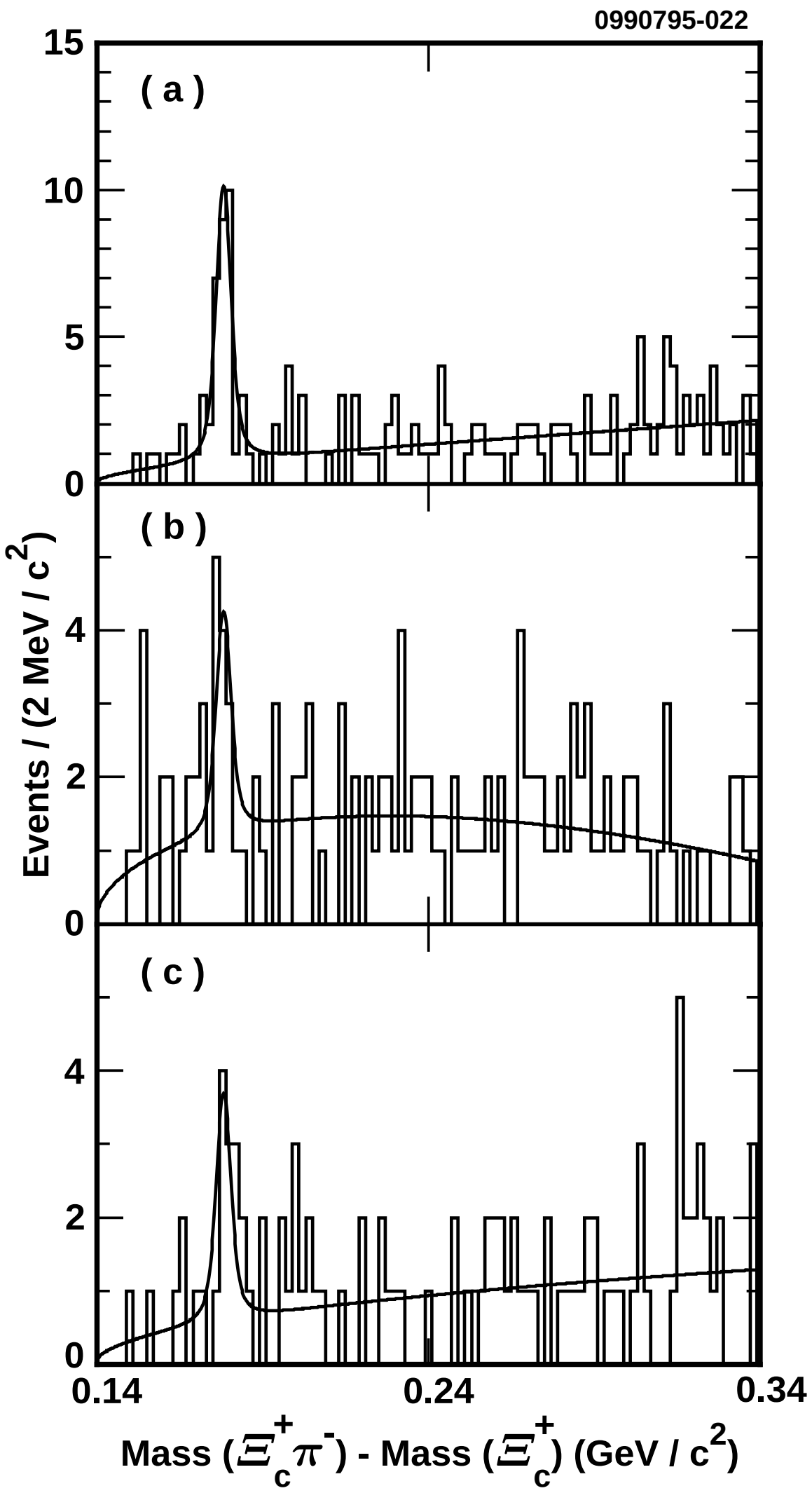

FIG. 3. The spectrum of the mass difference $M(\Xi_c^+\pi^-) - M(\Xi_c^+)$ for (a) Only $\Xi_c^+ \to \Xi^-\pi^+\pi^+$, (b) Only $\Xi^0\pi^+\pi^0$ and (c) Only $\Xi_c^+ \to \Sigma^+ K^{*0}$. The fits are described in the text.

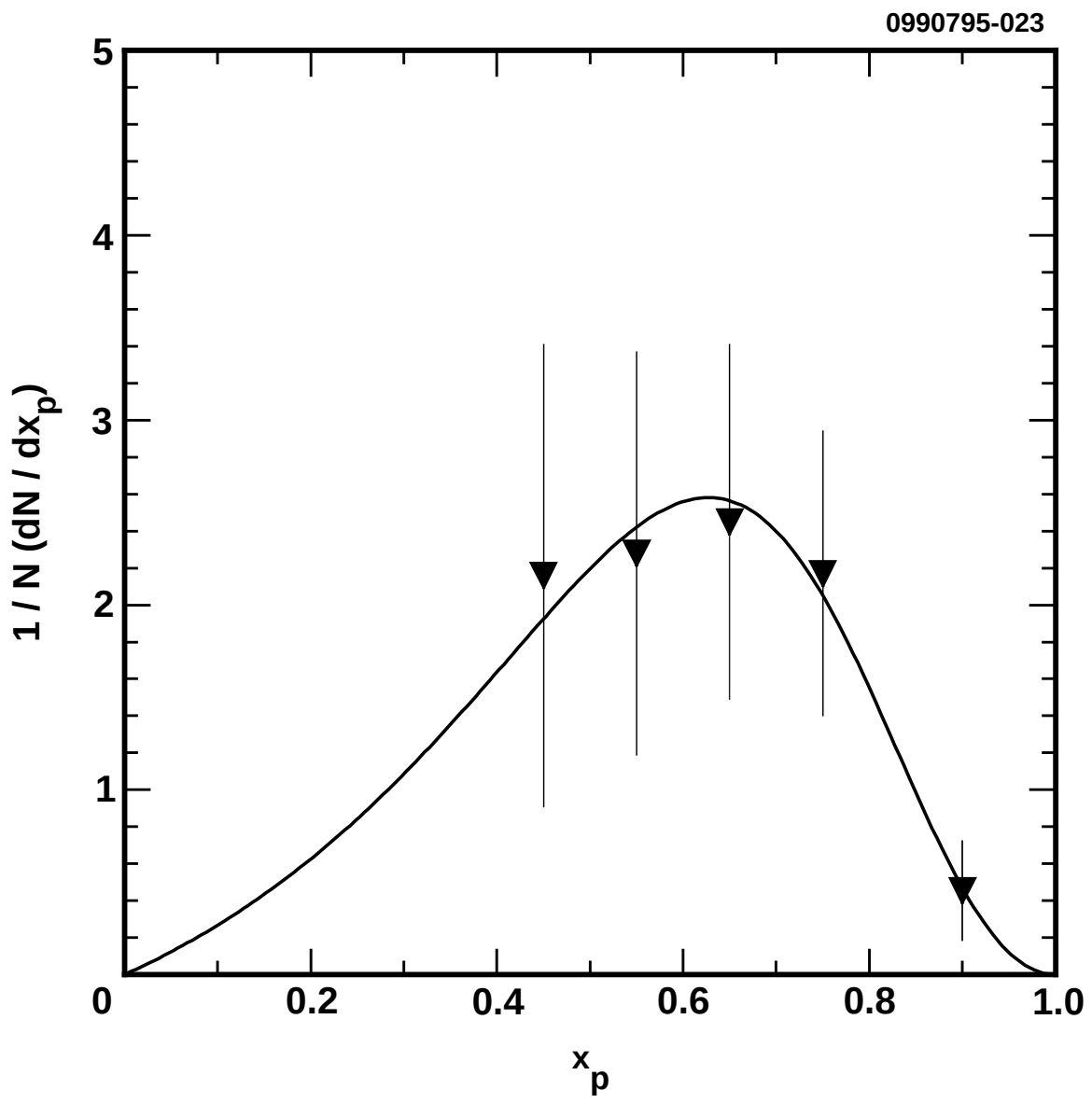

FIG. 4. The spectrum of scaled momentum, $x_p$, for the observed $\Xi_c^{*0}$ candidates. The fit is of the form of the Peterson function.

In conclusion, we have observed a narrow ($\Gamma < 5.5$ MeV/c$^2$ ) peak which we believe corresponds to the decay $\Xi_c^{*0} \to \Xi_c^+ \pi^-$. The mass difference $M(\Xi_c^{*0}) - M(\Xi_c^+)$ is measured to be $178.2 \pm 0.5 \pm 1.0$ MeV/c$^2$ .

## ACKNOWLEDGEMENTS

We gratefully acknowledge the effort of the CESR staff in providing us with excellent luminosity and running conditions. This work was supported by the National Science Foundation, the U.S. Department of Energy, the Heisenberg Foundation, the Alexander von Humboldt Stiftung, the Natural Sciences and Engineering Research Council of Canada, and the A.P. Sloan Foundation.